# Data to Decisions: *A Computational Framework to Identify skill requirements from Advertorial Data*

Aakash Singh, Anurag Kanaujia & Vivek Kumar Singh

**Abstract**

Among the factors of production, human capital or skilled manpower is the one that keeps evoving and adapts to changing conditions and resources. This adaptability makes human capital the most crucial factor in ensuring a sustainable growth of industry/sector. As new technologies are developed and adopted, the new generations are required to acquire skills in newer technologies in order to be employable. At the same time professionals are required to upskill and reskill themselves to remain relevant in the industry. There is however no straightforward method to identify the skill needs of the industry at a given point of time. Therefore, this paper proposes a data to decision framework that can successfully identify the desired skill set in a given area by analysing the advertorial data collected from popular online job portals and supplied as input to the framework. The proposed framework uses techniques of statistical analysis, data mining and natural language processing for the purpose. The applicability of the framework is demonstrated on CS&IT job advertisement data from India. The analytical results not only provide useful insights about current state of skill needs in CS&IT industry but also provide practical implications to prospective job applicants, training agencies, and institutions of higher education & professional training.

1. **Introduction**

The analysis of the availability and requirement of skilled human resources is an exercise that has been undertaken since the mid-1900s (Ivar berg, 1970), across multiple domains and by using varied approaches. As new technologies are developed and adopted, the new generations are required to acquire skills in these technologies in order to be employable. Multiple job portals showing job availability specific to the skill set of individuals (LinkedIn, Monster etc.) are available for both the employers and the employees. Some of these cater to specific industries/ sectors while some are general. Many organisations and consulting firms have also come up with their own analysis of skill requirements in selected target areas. For instance, a report from Deloitte looks at the skill gap in the manufacturing sector in the United States (Deloitte 2018), and a market research report released by British Council India focuses on skill requirements among young professionals in India (British Council, 2017), in selected sectors. These studies are important because of the economic, social, and political impacts that job availability in any economy can have on any society. Scholars have further explored changes in job markets and skill requirements over time. It has been observed that changing socio-economic and technological landscapes have a profound impact on the skill requirements.

Recently, the area of Computer Science & Information Technology (CS&IT) has evolved and led to a rapidly increasing role of technologies like AI, ML, Internet of Things (IoT), distributed computing, etc. across different sectors. This has made CS&IT as one of the most popular disciplines among the new generations. In fact, it has been postulated that it is relatively simple for students with computer science degrees to jump into this new wave of technology (West and Allen, 2018). Having expertise in CS&IT opens plenty of opportunities in terms of jobs, however, there is no systematic method to identify the prevailing skill needs and desired skill sets. Moreover, given the fact that CS&IT is a rapidly changing field, the desired skill sets for jobs also keep on changing. Therefore, it is important to identify required sets of skills that are in demand for different job roles in CS&IT industry.

Motivated by the need for identifying skill needs and its usefulness, the paper proposes a framework which takes advertorial data as input and produces useful analysis on required skills, connected skillsets, and training needs of applicants. The analytical framework uses a combination of techniques such as word2vec, affinitive propagation clustering, market-basket analysis etc. This proposed framework, can utilize data collected from online or offline digital database of job advertisements/availability. It processes this data to extract information using multi-level analysis. First content/statistical analysis is done to identify all the categories of job roles, industry requirements, average experience sought, educational requirements, locations of jobs etc. After this, skill clustering to group similar skills and observe patterns in different dimensions. These findings are used to provide detailed insights and recommendations for the related sectors. The developed framework has been tested on data from India for CS&IT jobs and the results are presented in terms of geospatial analysis, content analysis, and recommendation analysis.

The major contribution of this work is a Data to Decisions analytical framework that can produce useful analytical outcomes from advertorial data in an area. The framework is capable of both, descriptive as well as prescriptive analysis. The analytical results provide practical implications to prospective job applicants, training agencies, and institutions of higher education & professional training. The rest of the paper is organized as follows: Section 2 discusses some previous related studies in the area. The computational framework developed is described in section 3. Section 4 details the data collection, attributes, and data pre-processing. Analytical results are presented in section 5 followed by a Discussion of the results and suitability of the framework in section 6. The paper concludes in section 7 with a summary of the work, its importance, usefulness, and the practical implications.

## 2. Related Work

Most of the studies on the analysis of jobs and job skill requirements until recently had focused on the US and the European economies. These studies can be traced back to the 1960s, and they have looked at the changing job requirements and human capital available through the theoretical perspective of the Neo-classical view of labour markets (Horowitz and Herrnstadt, 1966; Scoville, 1966). Early studies defined the utility of such types of analysis for both economy and academia (Ivar berg, 1970). They used job vacancies and skills data from U.S. employment services sources to evaluate the changing skill requirements in the American economy (Rumberger, 1981).

Studies from the US and Europe have used interviews with managers and hiring personnel in software companies to identify skill deficiencies among the freshly graduating students in order to assist educators by suggesting areas where the curriculum in universities could be focused upon for improving employability (Radermacher et al., 2014). In the area of Big Data, the common skills of hired data scientists have been identified using a consensus-based pile-sorting method[1] (Gardiner et al., 2017). Similarly, content analysis of online job postings in Business Data Analytics and Data Science yielded a ranked list of relevant skills in specific categories. Organisational skills (decision making, organization, communication, and structured data management) were key to all job categories, and technical skills (like statistics and programming skills) were in most demand for data analytics (Radovilsky et al., 2018). In the case of Science, Technology, Engineering, and Mathematics (STEM) jobs, an analysis of changing basket of skills with time and technological change are documented (Deming and Noray, 2018). Professional skills required of accountants by employers in Australia and New Zealand were studied using content analysis of job advertisements (Tan and Laswad, 2018). While the skills required for postions of social media manager and social media marketer, both entry level jobs in U.S. were studied using data from indeed.com and employing independent raters to identify keywords and job skills from the data (Verma et al. 2021).

Studies based on online job postings in past couple of years are mostly focused on skills in the areas related to Information Communication Technology (ICT) and AI/ML, mostly in United States (Beblavy et al. 2016 and Acemoglu et al. 2021). More recent studies utilise web-based sources for acquiring such job and skill data. Using Job advertisement data on online job portals, some recent studies have identified skill sets sought after by employers for selected positions in Business and Data Analytics (Verma et al., 2019), and AI and ML (Verma et al., 2022). Some have applied a web content mining approach on portals such as Mosnter.com, Simplyhired.com etc. to extract computing jobs in the US and identified job categories and the associated skills need prevalent in the computing professions (Liteke et al., 2010). A recent study looked at over 2.3 million online job postings in Poland a national job posting aggregator SOJO (system of online job offers) which utilises web crawing methods to collect data from six national job sites (Arendt et al. 2023). Few other studies have identified the high and low-valued skills and traits in an entry-level IT worker (Aasheim et al., 2012; 2009a; 2009b). Even the application of ChatGPT as a skil and skilling tool has been explored (Opara et al. 2024).

---

[1] Consensus based pile sorting involves codification and identification of jobs and skill categories in the advertisement data by domain experts.

Recent studies like these have explored the correlation between job availabilities and relevant skill requirements in different areas of work. The findings from these studies are intended to supplement both the relevant industry by helping in designing better-curated job vacancies, and the academia by identifying skills which should be imparted to the students. However, most of these studies are conducted for the US and European regions. Such studies are not available in the case of developing countries such as India or Brazil. Additionally, these studies present limited information such as advertisements for specific jobs, skills in demand etc. which is largely in line with the traditional statistical analysis.

The present study aims to address this gap by presenting a framework that takes advertorial data as input and identifies broad range of relevant skills required, including connectedness of different skills. The advertorial data for job openings in the CS&IT sector of India is used as an example data to demonstrate the framework. To the best our knowledge, the study is the first of its kind to identify a broad range of computer science skills required in CS&IT area. Accordingly, the prospective applicants and job seekers could know about and learn the specific skills required. The set of skills connected with each other are also identified and are presented as recommendation to user, which can help them in improving their suitability for various job roles. In addition, the findings will be further useful for institutions of higher education and professional training in improving their academic curriculum by incorporating courses focused on developing the identified skills and skill sets.

### 3. Framework for Analysis of Job Skills

This study is a shade of data analytics where the job advertisement details scraped from World Wide Web (WWW) served as the data. The study aims to produce a guiding framework for such kind of analysis. It involves 3 steps, data extraction, data pre-processing, and data analysis. The primary focus of the study revolves around the last 2 steps i.e., data pre-processing and analysis and hence are discussed in detail. However, the first step is also crucial to the process hence a brief description of it is also accompanied.

### 3.1 Data Extraction

The data required are present on the open internet hence, extraction encompasses choosing an apt web scrapping library for crawling and parsing the WWW. The study recommends using Python's Srcapy framework for its structured approach. However, it is not binding and any similar tool will also meet the purpose. Once the scraping tool is ready the next thing desired is a target URL/website (Job portals,..) for which we want to conduct the study. Along with the URL, a key phrase is also required to start the extraction process.

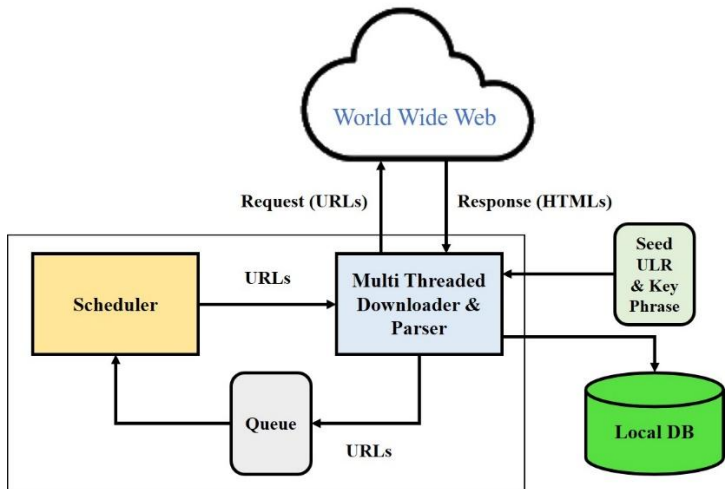

Figure 1: Data extraction process block diagram

This key phrase is analogous to the topic/field/specialization in which the study of job dynamics is required. The structure and formation of key phrases may differ for different websites being used however the intention remains the same. The process initiates with the downloader of the scraper starting a thread to send the request to the URL's web application and receive a response. The responses received can either be in the form of a new HTML or a Web API's HTTP response header. It is now the job of a custom Parser unit to decode and extract the required information from the response. In the first run, the downloader crawls all the hyperlinked pages containing job advertisements until null is encountered. In the meantime, the parser extracts the individual job advertisement URLs from the response and appends them to a Queue. The queue is then passed to a scheduler unit whose job is to assign one URL to one thread of the downloader and parser unit. To reduce the request load on the host website the number of simulations active threads should be restricted along with the maximum amount of requests per thread per sec. The aim of the parser unit in this run is to extract specified information from the job description page in the form of JSON documents which are later stored in a local collection. A representational block diagram for the extraction process is depicted in Figure 1.

### 3.2 Data Pre-processing

The data extracted in the previous section need to be pre-processed before being fed to any data analytics algorithm. A standard process involves cleaning data and making it free from any spurious terms or phrases. Apart from these, the study proposes 2 supplementary stages. First is the Semantic grouping of job advertisements based on fields (Job Role/ Name/ Title) having free user text and second is a clustering of skills in demand based on their appearance in similar advertisements. This additional pre-processing is introduced to further improve and generalize the analysis in later stages.

#### 3.2.1 Semantic Grouping

The fields present in the data may either belong to one of the 2 formats, standard tag format or free text format. The first format includes the fields which have fixed and defined tags given by the corresponding websites and hence offer a minimal challenge in the analysis phase. But, the second type (free-text format) may have much similar content that would have been characterized as different because of the different writing styles of different advertisers. There may also arise some typographical and grammatical errors due to its free text nature. Considering these deviants as separate types of advertisement will fail any count-based analysis. To meet the challenge, the study proposes to employ Natural Language Processing (NLP) techniques. The first step starts with the removal of standard stop-words using Python's NLTK library. Later, a deep learning-based Word2Vec technique is proposed to find the semantic similarity among free-text type fields. To add Google's pre-trained "Google News 300" Word2Vec model, which is trained on 300 billion google news words is suggested to find out the word mover's distance (WMD) (Kusner *et al.* 2015) between each pair of individual fields. Once an (n * n) distance values matrix is ready, where n is the number of unique entries of the field. Application of affinity propagation-based clustering to group semantically similar values of field is proposed. This would result in the formulation of clusters of values having semantic similarities. After the cluster formulation, the value with max number of appearances in the field's cluster is appointed as the leader of the cluster. All further counting is proposed in the way that if any element of a cluster is encountered, its leader's count should be increased.

#### 3.2.2 Skill Clustering

Skills in demand are among the major parameters that any such advertorial analysis tries to evaluate. The aim of the step is to group skills based on their occurrence/role similarity and classify them in a broader class (Figure 2). This indeed will extend the horizon of the study to capture a much wider snap from the available data.

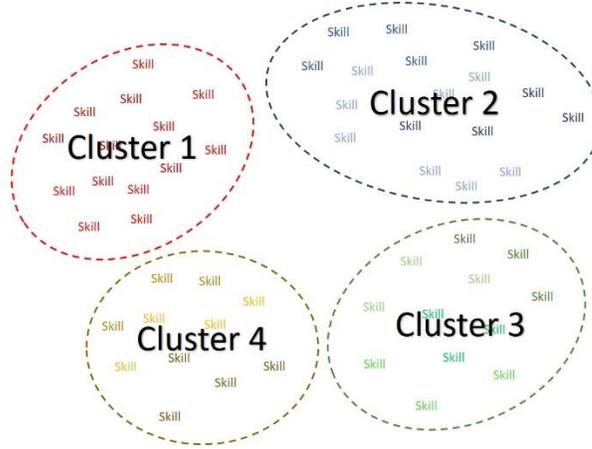

Figure 2: Role-based skill clusters (desired output of the step)

While similar types of studies earlier (Verma *et al.* 2019, Verma *et al.* 2022) have followed a manual annotation process to segregate skills into their respective skill cluster, contrary this study proposes an automated Cosine similarity-based clustering technique to extract similar and meaningful clusters out of the given skill list. If the data contains lots of unique skill tags it is recommended to limit the skills based on a threshold of some minimum number of appearances in unique advertisement. This will aid in reducing the computation intensity of all further steps to be followed along with reducing any noise that has crawled in with less common skills.

For the computation of skill cluster, the study proposes the construction of a job-skill matrix, as,

$$M_{js} = \begin{matrix} a_{00} & a_{01} & ... & a_{0m} \\ a_{10} & a_{11} & ... & a_{1m} \\ \vdots & \vdots & \ddots & \vdots \\ a_{n0} & a_{n1} & ... & a_{nm} \end{matrix}$$

Where,  n = total number of jobs
and,  m = total number of unique skills

In the matrix, $M_{js}$, a row corresponds to a job and a column corresponds to respective skills. The next step is to normalize this matrix. For this division of each row with its L2 norm is suggested. The L2 norm is calculated using (i).

$$\sqrt{\sum_{j=0}^{k}(a_{ij})^{\wedge}2} \qquad ...(i)$$

where $a_{ij}$ is the element in matrix $M_{js}$ at $i^{th}$ row and $j^{th}$ column.

Let the normalized job-skill matrix be called as $N_{js}$. To find the Cosine Similarity between all elements (skills) in the matrix $M_{js}$, it is recommended to take the dot product of $N_{js}^T$ ($N_{js}$ transpose) and $N_{js}$. The resultant matrix $M_{ss} = (N_{js}^T \cdot N_{js})$ would provide us with the desired skill-skill Cosine Similarity (Alag 2008). With the preparation of the matrix $M_{ss}$, the next step is to extract clusters out of this new information that it carries. The use of VOSviewer to run the clustering algorithm is recommended here. The step is expected to produce groups/ clusters of skills having role-based similarity (Figure 2).

### 3.3 Data Analysis

This step involves the actual conversion of the data processed to some meaningful information. The study proposes the use of 3 major types of analysis to meet the purpose. They are content analysis (Neuendorf, 2017), geospatial analysis (Smith *et al.* 2007), and recommendation analysis (Agrawal *et al.* 1994). A brief description of the structure and working of each of them is as followed

3.3.1 ***Content Analysis***: It refers to a type of quantitative as well as qualitative analysis whose sole purpose is to mine trends for the given data. It includes the study of documents/texts of various formats to examine patterns in communication in a replicable and systematic manner. The study recommends the use of the frequency count method on important parameters existing in the processed corpus.

3.3.2 ***Geo-Spatial analysis***: The objective of this analysis is to identify the physical locations where job opportunities were available and other attributes associated with the job market. This includes the extraction of location information from the advertisement fields. Followed by the conversion of that textual information to its approximate longitudinal and latitudinal. values. Pythons GeoPy library is suggested for this task. To add, the GeoPandas library is recommended for plotting the various correlated parameters.

3.3.3 ***Recommendation Analysis***: In addition, a recommendation analysis is also been suggested to identify context-based skill sets which would otherwise be non-obvious. A recommendation analysis is a form of modern data mining/machine learning technique, which perform information filtering to provide suggestions for possible choices out of available options aiding in decision-making. This study recommends content-based filtering to find appropriate skills to match existing sets. This analysis is similar to what most technology giant e-Commerce companies employ nowadays, i.e., market basket analysis (MBA). This will help to reveal several context-based hidden skill sets. These skill sets could prove to be extremely helpful to an individual who works or aspires to work in a particular industry. We propose Apriori algorithm to identify frequently occurring skill pairs and their recommendation rules derived from them. The lift is the recommended parameter for selecting the relevant recommendations. To understand let X and Y be two co-occurring skills or skill sets, then the lift value K for X → Y tells us that the probability of an advertisement asking for skill set Y increases by the value of K provided that skill set X has already been asked in the advertisement. Mathematically it is defined as

$$Lift\ (X \rightarrow Y) = \frac{Confidence\ (X \rightarrow Y)}{Support\ (Y)}$$

Where,

$$Confidence\ (X \rightarrow Y) = \frac{Support\ (X \cap Y)}{Support\ (X)}$$

and,

$$Support\ (X \cap Y) = \frac{No.\ of\ advt.\ containing\ both\ skill\ set\ X\ and\ Y}{Total\ no.\ of\ advt.}$$

$$Support\ (X) = \frac{No.\ of\ advt.\ conatining\ skill\ set\ X}{Total\ no.\ of\ advt.}$$

$$Support\ (Y) = \frac{No.\ of\ advt.\ conatining\ skill\ set\ Y}{Total\ no.\ of\ advt.}$$

## 4. Results and Validation

In order to test the applicability of the proposed framework, job postings in "Computer Science" and "Information Technology" on the Indian job postings website, Naukari.com was extracted and analysed. The following section describes the process and results from the analysis (Figure 3).

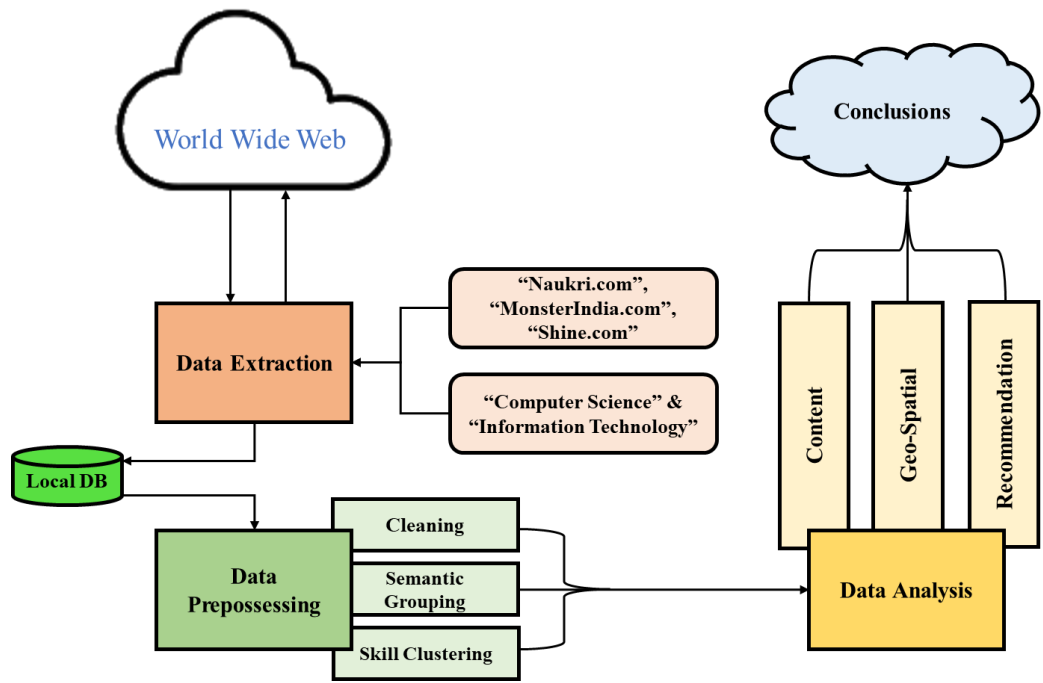

Figure 3: Framework validation block diagram

**4.1. Selection of Data Source:** Data from three popular websites namely, MonsterIndia.com[2], Naukri.com[3] and Shine.com[4], were scraped and compared. The data from Naukri.com was used for analysis due to a few decisive reasons. First, Naukri.com is among one of the most popular websites in the world based on net global traffic with about 28 million[5] unique visitors being directed towards the website in a month. Also, in India, it ranks 2nd in website category ranking (Jobs and Employment) according to a well know website traffic analysis company named SimilarWeb. Second, the content of open data available on the website was found to be the most extensive among others.

**4.2. Data Layout:** The data was extracted from the online portal using Scrapy, in August 2021. The structure of the data is summarized in **Table 1**. A total of 30,807 unique job advertisement records were successfully fetched and stored for our examination purposes. The majority of the advertisements (~64%) were found to be of the year 2021, year wise distribution is shown in **figure 4**. This askew nature of the data makes it a bit unsuitable for any kind of temporal analysis.

---

[2] Monster India (Estd. 2001) is India's leading online career and recruitment resource. It is a part of Monster.com, which is a global online employment solution, functional for more than 20 years.
[3] Naukri.com is an Indian employment website operating in India and Middle East. It was founded in March 1997. Naukri.com is the largest employment website in India.
[4] Shine.com is the fastest growing Online Job Portal in India. It was launched in 2008, and it has more than 3.86 crore registered users, over 80 lakh active users and more than 3 lakh advertised job vacancies.
[5] https://www.similarweb.com/website/naukri.com/#traffic

**Table 1.** Data fields and their types in data scrapped from Naukri.com job requirement data.

| Fields | Type |
|---|---|
| Job name | Text |
| Company Name | Text |
| Advertisement date | Date |
| Apply count | Int (Integer) |
| View count | Int |
| Role category | Text |
| Education | List |
| Industry | Text |
| Minimum experience | Int |
| Maximum experience | Int |
| Employment type | Text |
| Functional area | Text |
| Locations | List |
| Key skills | List |
| Vacancy | Int |
| Salary | List |
| Description | Text |
| Ambitionbox-details | List of lists |

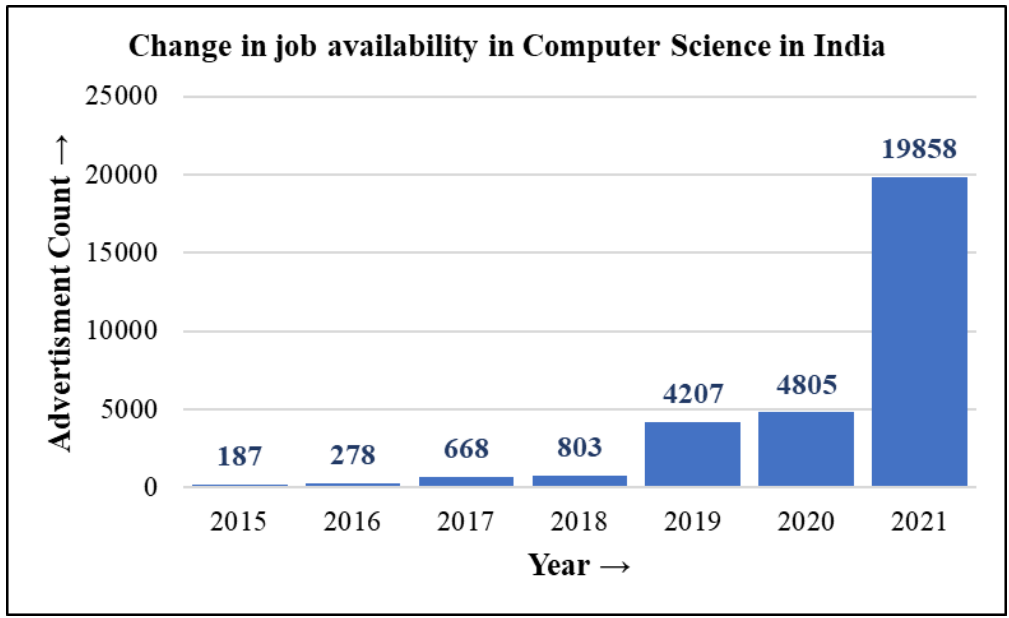

**Figure 4.** Numbers of published advertisements for CS jobs retrieved from Naurkri.com

**4.3. Data Pre-processing:** The fields that were in free text format were processed using Natural Language Processing (NLP) techniques. First, we have applied the standard stop word removal method available to us by python's NLTK library. Later, the deep learning-based *Word2Vec* technique was used to find the semantic similarity among job titles. We have used Google's pre-trained "Google News 300" *Word2Vec model*, which is trained on 300 billion google news words to find out the word mover's distance (WMD) (Kusner el at. 2015) between each pair of individual job titles. Once we were ready with (n * n) distance values, where n is the number of unique job names, we have applied affinity propagation-based clustering to group job names having semantic similarity. A total of 165 groups were formulated. After the clusters were formed, we assigned the job name with max number of appearances as its leader. All later counting was done in the way that if any element of a cluster was encountered, its leader's count was increased.

**4.4. Data Analysis:** Initially we had a total of 5,228 unique skill names in our dataset. To filter out unnecessary noise from skills asked less frequently, and to reduce computation intensity, we have only considered those skill that have appeared in 20 or more job advertisements. This step reduced the total skills for consideration to 1,063. For computation of skill similarity we have created a job-skill matrix. Following this the matrix $M_{ss}$ was prepared, and skill clusters were extracted. We have used VOSviewer to run clustering algorithm (Waltman et al., 2010) and later plotted the discovered cluster using the same. A total of 11 distinct skill cluster could be identified **(Figure 5)**.

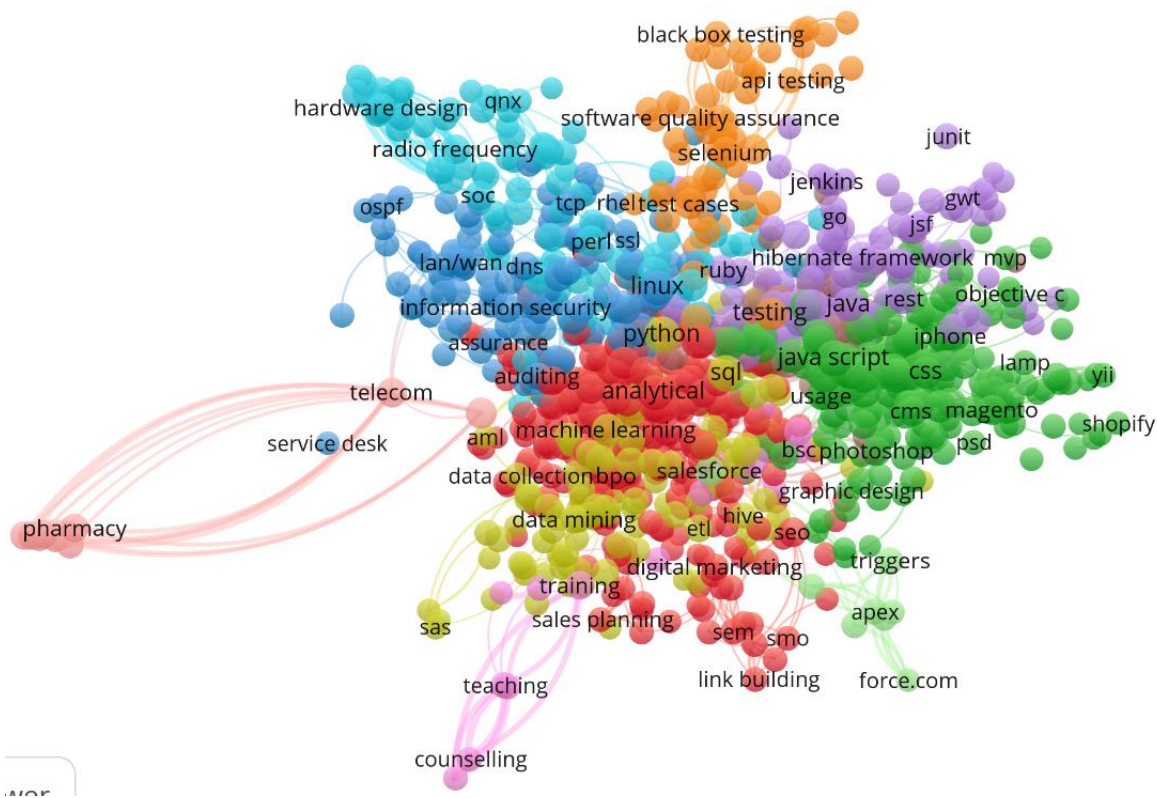

**Figure 5.** Mapping of the different types of skills into clusters using VoS Viewer

We have tried to provide suitable names to these clusters based on skills that they contained. We consulted with some independent raters (university faculty and industry professionals) to provide their expertise by agreeing with the given name or suggesting an alternative name for the skill clusters. **Table 2** contains the most favoured name of the clusters and their respective sample skills. These clusters were utilised in our analysis to visualise a macro level skill set requirements in various scenarios.

**Table 2.** Classification of skills in clusters of different application domains

| Cluster | Constituent Sample Skills |
|---|---|
| Product management | Analytical, agile, project management, monitoring, scrum, product management, work flow, erp implementation, customer service, risk management, digital marketing, time management etc. |
| Web Development | java script, mysql, html, php, jquery, web technologies, android, xml, ajax, front end, json, mvc, wordpress, etc. |
| Systems Support | Linux, automation, trouble shooting, windows, networking, unix, technical support, http, configuration management, virtualization, information security, shell scripting, vmware, etc. |
| Data Science | Python, sql, machine learning, performance tuning, data analysis, business intelligence, data management, data modelling, artificial intelligence, database design, big data, etc. |
| Application development-Java Technologies (I) | Java, data structure, git, software development, web service, aws, core java, hibernate framework, nosql, rdbms, j2ee technologies, mongodb, apache, rest, angularjs, react.js, etc. |
| Embedded systems | Debugging, cpp, perl, software designing, graphic, soc, wireless, simulation, automotive, middleware, firmware, radio frequency, system architecture, system integration, etc. |
| Software Testing | Sdlc, testing, test cases, jira, selenium, automation testing, performance testing, testing tools, manual testing, functional testing, software testing, test scripts, test planning, system testing, etc. |
| Application development-.Net framework (II) | Oracle, application development, Microsoft, .net, sql server, c#, asp dot net, soa, wcf, object oriented design, visual studio, six sigma, wpf, asp.net mvc, visual basic, asp, linq, vb net, etc. |
| Education & Training | Training, teaching, mentoring, counselling, advisory, engineering, professor activities, etc |
| IT applications | health care, telecom, pharmacy, physical education, hotel management, business studies, botany etc |
| CRM-Salesforce | Salesforce, apex, salesforce.com, visual force, triggers, sfdc, force.com, etc |

## 2.4 Content Analysis

### 2.4.1 *Top Job roles*

We investigated the popular job roles offered by the CS&IT recruiters. For this purpose, document distance metric called Word Mover's Distance was employed using the NLP tool 'Word2vec'. This tool creates word embeddings using a collection of linked models which are shallow, two-layer neural networks trained to reconstruct linguistic contexts of words. This analysis resulted in the identification of top job roles based on their appearance in advertisements text (Table 3).

Table 3: Advertisement Count for different Job profiles in CS&IT

| Rank | Job | Advertisement Count |
|---|---|---|
| 1 | php developer / senior php developer | 4403 |
| 2 | android developer / senior android developer | 2190 |
| 3 | software engineer / senior software engineer | 1238 |
| 4 | software architect | 454 |
| 5 | mobile app developer | 423 |
| 6 | front end developer | 411 |
| 7 | python developer | 408 |
| 8 | full stack php developer | 404 |
| 9 | ios developer | 389 |
| 10 | data scientist / senior data scientist | 657 |
| 11 | support engineer | 336 |
| 12 | java developer / senior java developer | 659 |
| 13 | full stack developer | 327 |
| 14 | dot net developer | 326 |
| 15 | technical project manager | 325 |
| 16 | salesforce developer | 324 |
| 17 | qa automation engineer | 320 |
| 18 | senior data engineer | 315 |
| 19 | react.js developer | 305 |
| 20 | software test engineer | 293 |

The list is a mixture of area specific and general job roles relating to CS&IT, with the former being dominated by Php developers contributing about 14% (Php developer + Senior Php developer) of total available job roles in our dataset. Apart from that significant demand of Android developers can also be observed. In later category software engineer and software architect were seen as most popular job roles.

### *2.4.2* Top skills in demand

The individuals with expertise in java script, python, sql, mysql and php were the most sought after by employers (Table 4). The top five skills contribute a total of 34% in the total of the 1063 skills identified related to the available jobs. This observation is in line with the available job roles where php and android developers are most sought after by the computer science firms.

Table 4: Top Skills based on requirements of Skills in different Job Advertisements

| Rank | Skill | Occurrence Count |
|---|---|---|
| 1 | Java script | 4783 |
| 2 | Python | 3687 |
| 3 | Sql | 3531 |
| 4 | Mysql | 3515 |
| 5 | Php | 3506 |
| 6 | Html | 3240 |
| 7 | Jquery | 2737 |
| 8 | Agile | 2658 |
| 9 | Linux | 2650 |
| 10 | Coding | 2469 |
| 11 | Xml | 1838 |
| 12 | Ajax | 1804 |
| 13 | Automation | 1753 |
| 14 | Android | 1633 |
| 15 | Web technologies | 1559 |
| 16 | Debugging | 1487 |
| 17 | Front end | 1449 |
| 18 | Json | 1323 |
| 19 | Data structure | 1268 |
| 20 | Open source | 1268 |
| 21 | Mvc | 1245 |
| 22 | Cpp | 1244 |
| 23 | Wordpress | 1167 |
| 24 | Git | 1161 |
| 25 | Windows | 1147 |
| 26 | Networking | 1081 |

### *2.4.3 Top industries offering CS&IT related jobs*

Among the industries which were looking for personnel in CS&IT jobs, IT-Software and Software Services (ITSS) industry accounts for more than 67 percent job advertisements. The rest of the 32 percent advertisements are from varied industries ranging from Recruitment, Electronics, e-commerce, Education, Media and Telecom sector (Table 5).

Table 5: Ranking industries based on the number of jobs available in CS&IT

| Rank | Top Industry | Count |
|---|---|---|
| 1 | IT-Software, Software Services | 17892 |
| 2 | Recruitment, Staffing | 2617 |
| 3 | Semiconductors, Electronics | 976 |
| 4 | Internet, Ecommerce | 919 |
| 5 | Banking, Financial Services, Broking | 875 |
| 6 | Strategy, Management Consulting Firms | 824 |

| 7 | Education, Teaching, Training | 743 |
| 8 | BPO, Call Centre, IteS | 691 |
| 9 | Media, Entertainment, Internet | 618 |
| 10 | Telcom, ISP | 493 |

A detailed breakup of the top 5 skill clusters in demand and the respective top 5 skills for these clusters was prepared and visualised using a sunburst chart (figure 4). The circle at the center shows percentage contribution of advertisements, inner ring shows the name of the industry, middle ring shows the skill cluster and other outer ring shows the top skills. Two charts were prepared and color coded for clarity as one industry has very large number of advertisements. The first chart (top) shows top skill clusters and their respective top skills for IT-Software, Software Services. Similarly, the second chart (bottom) shows top skill clusters and their respective top skills for the other four Industries. Web Development, Data Science, Project Management, Application Development – I (Java Technologies), and System Support are the top skill clusters in ITSS. In Recruitment, Staffing, the top five skill clusters are Web Development, Product Management, Data Science, Application Development – I, and System Support. Some of these skills are common in other three sectors as well, indicating the high level of overlap that exists across these areas in terms of skill requirements from computer science professionals.

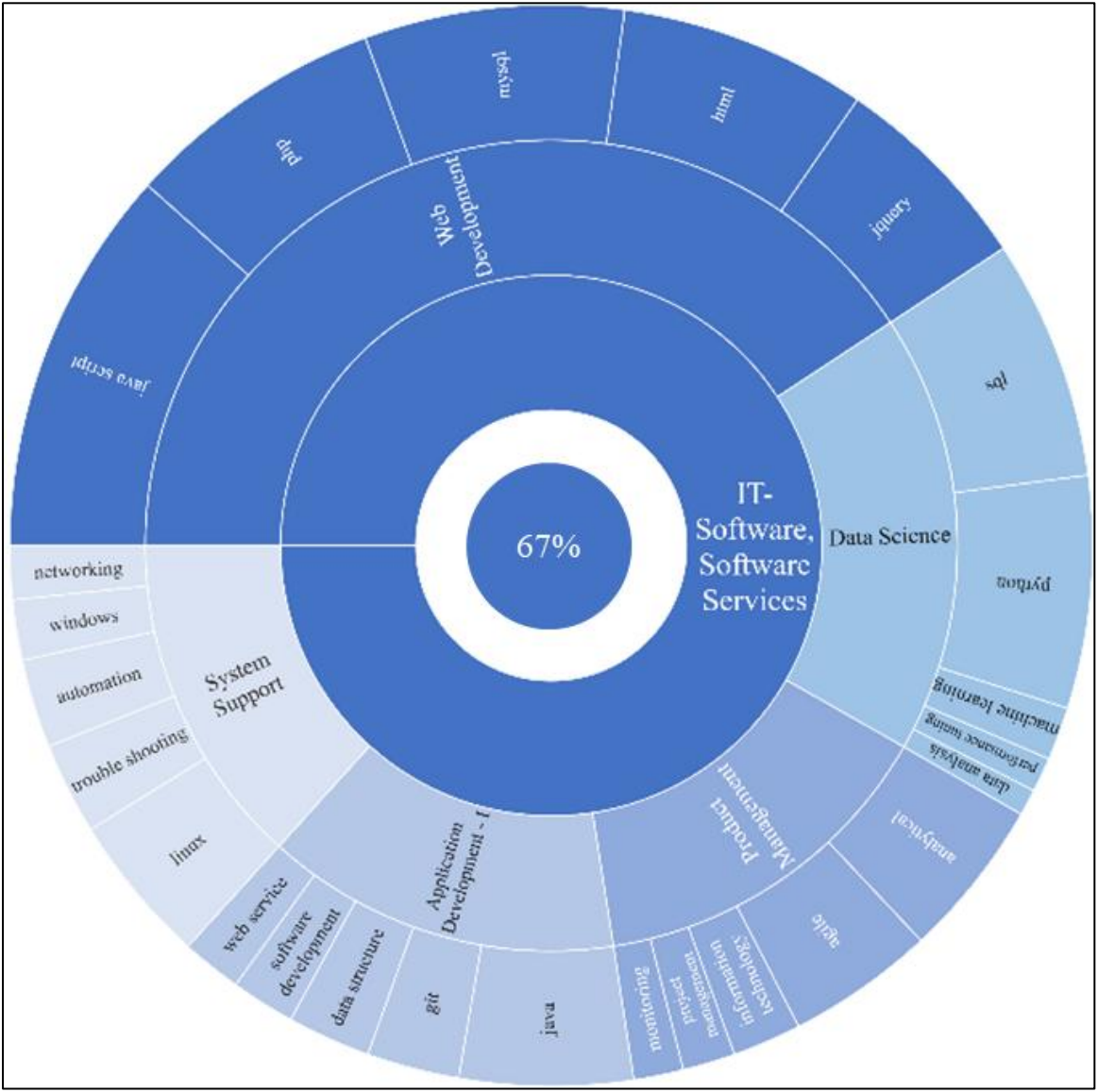

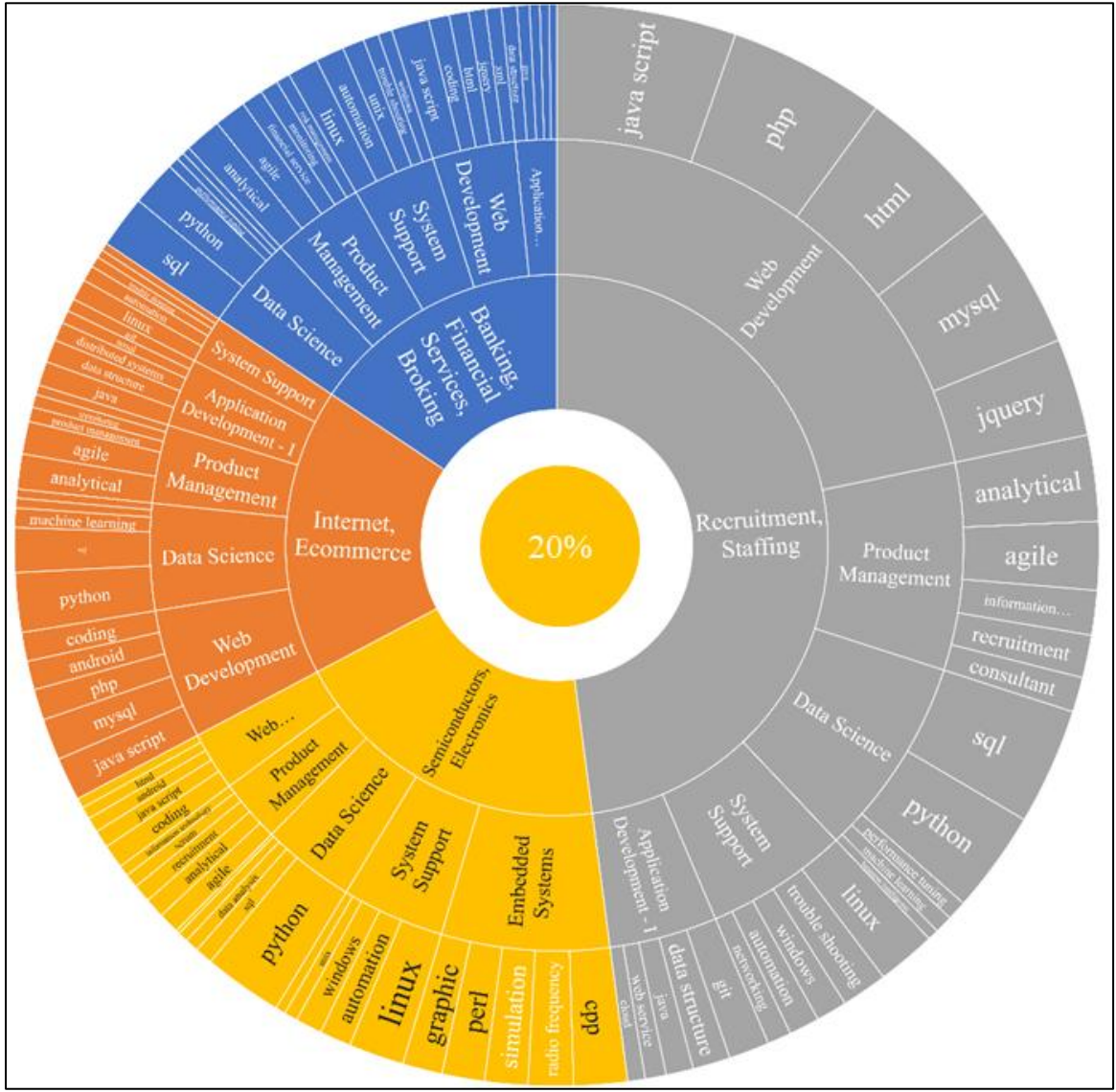

Figure 4: Most demanded skills in the Top 5 industries related to CS&IT in India contribute about 85% of the total jobs in the discipline. Rest 15% of jobs in CS&IT are related to other industries (not shown in the figure). (a) IT-Software and Software Service Industries account for more than 65% of available jobs in CS&IT. (b)Recruitment/Staffing, Semiconductors/Electronics, Internet/E-commerce, and Banking/Financial Services/Broking contribute to about 20% of the jobs available in CS&IT.

### *2.4.4 Freshers vs Experienced*

Freshers were those candidates who had no prior work experience when applying for the advertised job positions, the advertisements for these positions were compared with advertisements for positions where the required work experience was greater than the average work experience across all advertisements (i.e., 5.47 years). The findings for these were as follows.

Most of the recruiters were looking for web development professionals in case of freshers, with about 39 percent advertisements in this skill cluster. For the positions where experienced professionals were being sought, web development jobs constituted 22 percent of the total distribution followed by Product management, Application development-I, Data science and Systems support. Similar pattern could be seen for the freshers as well (Figure 5).

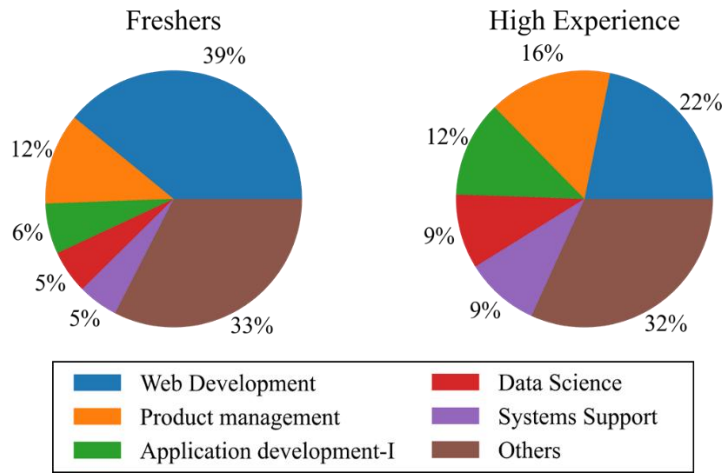

Figure 5: Distribution of skill cluster required between CS jobs available for Freshers and for Experienced professionals (with minimum 5.47 years of experience) in India.

### *2.4.5 High vacancy and High application*

Most of the high vacancy jobs were available for "Web development" professionals, with about 30 percent advertisements in this cluster followed by Application development-I (17 percent), Product Management (5 percent), Application development-II (4 percent), and Data Science (4 percent). For the positions which received high number of applications, we may observe a change in the distribution. High number of applications were submitted from professionals in response to vacancies in "Web development" (21 percent) followed by Product management (13 percent), Data Science, Application Development – I (11 percent) and System Support (8 percent). (Figure 6). The advertisements with high vacancies represent the market demand whereas those receiving high applications represent the available supply of the skilled professionals. Moreover, we observed a larger number of applications for jobs relating to the Data Science cluster (11 percent) depicting the growth and popularity of the domain. However, the low number of advertisements with high vacancies indicates a smaller demand for these skills in computer science related jobs. Thus, a gap is observed in volume of demand and supply according to skill sets.

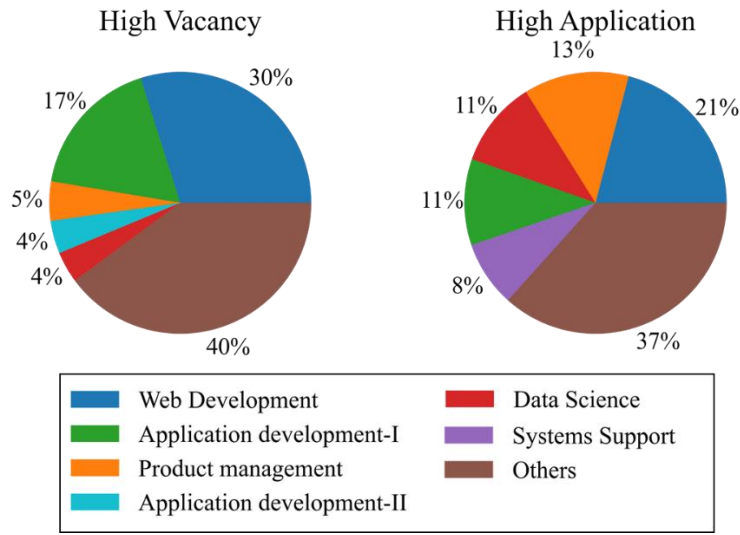

Figure 6: Distribution of skill cluster required between CS&IT jobs available for High vacancy and for high application receiving jobs in India.

The frequently occurring skill sets (with various cardinalities) were also traced. The support values calculated for each skill and group of skills were used to identify the most sought-after skills in jobs which advertised a high number of vacancies and the jobs which received a high number of applications per advertised position (Table 6). Among the two groups, some subtle but interesting differences were observed. Java Script and Java were the most sought-after skills in case of jobs which had high number of vacancies (more than 4), while python and sql were the most sought-after skills in case of jobs receiving high applications. Machine learning and python did not feature among the top five frequently occurring skills for high vacancy jobs. It can also be argued that the two groups represent skill demands and skill availability respectively.

Table 6: Top two frequently occurring (one, two, three and four) skills for high vacancy positions vis-à-vis positions receiving large number of applications. These were identified by their support values using Market basket analysis.

| Skills sought after for High Vacancy jobs | Support Value | Skills sought after in job receiving High Applications | Support Value |
| --- | --- | --- | --- |
| Java script | 0.2422 | Python | 0.1609 |
| Java | 0.1796 | Sql | 0.1480 |
| Html, css | 0.1251 | Html, java script | 0.0639 |
| Jquery, Java script | 0.0979 | Machine learning, Python | 0.0445 |
| Java script, html, css | 0.0897 | Html, java script, jquery | 0.0269 |
| Mysql, php, java script | 0.5850 | Mysql, php, java script | 0.0248 |
| Java script, jquery, html, css | 0.0448 | Jquery, mysql, php, java script | 0.0155 |
| Mysql, php, jquery, java script | 0.0380 | Css, html, java script, jquery | 0.0148 |

## 2.5 Geo-spatial Analysis

This section contains results from the geo-spatial analysis showing the top cities where jobs are located and the differentiation in openings for freshers and experienced professionals.

### 2.5.1 Top Cities for CS&IT related Jobs

Bangalore, Hyderabad, Pune, Mumbai, and Chennai are the top cities accounting for more than 70 percent of total advertised positions. The breakup of the different job clusters in these five cities is shown in figure 7. Following the overall trend, Web Development had the highest percentage of available jobs in all the cities. Bangalore with 9590 advertisements contributed to about 31 percent jobs, Hyderabad with 3661 advertisements contributed 11 percent, Pune with 3420 jobs contributed 11 percent, Mumbai with 2784 advertisements contributed 9 percent, and Chennai with 2274 advertisements contributed 7 percent jobs.

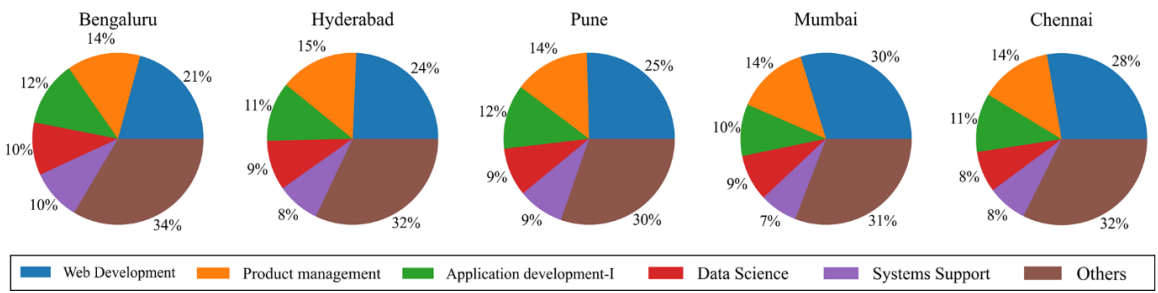

Figure 7: Distribution of available jobs in different regions of the country. Breakdown of specialisations of CS&IT Jobs in major Indian cities

### 2.5.2 Top cities with openings for freshers and experienced professionals

A Geo-spatial analysis showed that the regions near Delhi NCR and the northern parts of the country were the best suited to freshers, followed by Hyderabad and Mumbai regions. On the other hand, Bangalore and the southern parts of the country were the best suited for experienced professionals. There are almost no advertisements for freshers from the north-eastern states (Figure 8).

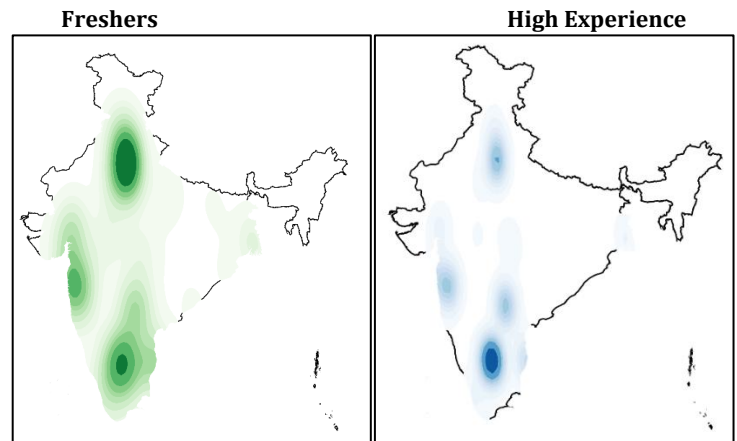

Figure 8: Distribution of CS&IT jobs for beginners (with no prior experience, green) and experienced (with minimum 5.47 years of experience, blue) in India

## 2.6 Recommendation Analysis

The recommendation/ MBA algorithm discussed earlier was used to obtain the most relevant recommendations for top 10 frequently demanded skills in advertisements (Table 6). The table displays three types of information, the skill (antecedent), corresponding recommendations (consequents), and its subsequent lift value. The inference of the table can be taken as follows: the chances of asking the consequent skills increase by "L" (lift) times when an advertiser has already asked for antecedent skills. We have only considered the top 2 relevant recommendations per antecedent skills, however more such recommendations were fetched using the algorithm. Among job vacancies for freshers, the top five antecedent skills with the highest support values were observed included Javascript, Php, HTML, Mysql and Jquery. In case of vacancies for experienced professionals, the top five antecedent skills were Python, Java Script, sql, Agile and Linux (Table 7). This relationship between antecedent and consequent skills can form recommendation for the stakeholders namely, professionals and course instructors. For instance, if a fresh graduate has skills related to php, s/he may look to build skills in web technologies, msql, ajax and jquery to have better opportunity of getting hired. The course designers and instructors can similarly use these recommendations to include relevant topics in the course and exclude skills not in demand. Similarly, experienced professionals can also utilise the recommendations for guiding their skill development efforts by acquiring relevant skills recommended.

Table 7: Top two frequently occurring (one, two, three and four) skills for high vacancy positions vis-à-vis positions receiving large number of applications. These were identified by their support values using Market basket analysis.

| **Freshers** (antecedent → {consequents} lift) | **Experienced** (antecedent → {consequents} lift) |
| --- | --- |
| Java script → {html5, php} 4.123 | Python → {machine learning} 3.266 |
| Java script → {web technologies, php, css, jquery} 3.865 | Python → {automation, linux} 3.092 |
| Php → {web technologies, msql, ajax, jquery} 4.147 | Java script → {jquery, css} 4.345 |
| Php → {html, css, development} 4.147 | Java script → {ajax, jquery, html} 4.303 |
| Html → {mysql, c} 5.384 | sql → {analytical, python} 2.480 |
| html → {java script, oracle, sql} 5.384 | sql → {oracle} 2.152 |
| Mysql → {wordpress, ajax, php, jquery} 5.824 | Agile → {scrum} 4.101 |
| Mysql → {cpp, c, html} 5.781 | Agile → {jira} 2.453 |
| Jquery → {web technologies, mysql, ajax, java script} 6.022 | Linux → {automation, python} 3.557 |
| Jquery → {php, json} 5.978 | Linux → {perl} 3.205 |
| (java script, html) → {.net, oracle, sql} 6.730 | Html → {java script, css} 6.186 |
| (java script, htm;) → {bootstrap, css} 5.811 | Html → {xml, jquery} 5.876 |
| Sql → {.net, java script, python} 8.061 | Analytical → {machine learning} 2.202 |
| sql → {oracle, html} 7.255 | Analytical → {sql, python} 2.014 |
| (Php, java script) → {web technologies, msql, ajax, jquery} 7.33 | Mysql → {ajax, php} 6.925 |
| (Php, java script) → {css, development} 6.871 | Mysql → {wordpress} 5.171 |
| Python → {oracle, sql} 5.890 | Java → {spring} 6.969 |
| Python → {machine learning} 5.867 | Java → {algorithm} 4.490 |
| Android → {sqlite} 5.309 | Coding → {debugging} 2.731 |
| Android → {xml, json} 3.887 | Coding → {data structure} 2.183 |

## 5. Discussion

The study presents a systematic analysis towards identification of skill needs from the advertorial data in the area of CS&IT. The advertisements are analysed in three respects: content, geospatial and recommendation analysis. A total of eleven clusters of skills were created by using affinity propagation-based clustering of job skills on extracted job advertisements. These clusters represent areas of activity in CS&IT in India and provide the first clues into a comprehensive picture of the national job market. The most number of advertisements were from the IT & business services industry which accounts for approx. 8% of the overall GDP of India and is expected to grow three times its current size to up to US$ 19.93 billion by 2025 (IBEF, 2022). This is coupled with the observation that India is among the topmost offshoring destination for IT companies worldwide indicate an upcoming demand for skilled individuals from this area. As per the analysis, the most sought-after skills were in Web Development, Data Science, Project Management, Application Support-I (Java technologies), System Support, and Embedded Systems. These are the areas where major growth would be expected to happen. The growth of startups in edutech (Byju's, Unacademy), food and groceries delivery (Swiggy, Liscious, Big Basket), Debt marketplace (Credavenue), data services (Hasura, Amazon Web Services), fintech etc. with record investments (US$ 36 billion in 2021 alone) could hold a causal relationship with the high demand of professionals with the most sought-after skills. The observation that more than half (69%) of the advertised job openings were for the top five cities with Bangalore having approximately one-third (31%) of the total advertisements is another interesting aspect to note in terms of territorial mapping of job availability. The ongoing push towards online services (Digital India) both at government and at private industry level could be a contributing factor for the high demand of skills in 'web development'. It may also be noted that most number of advertisements came from this cluster.

Freshers had a higher number of opportunities in Delhi NCR and northern parts of the country if they possessed skills in Java script, PHP, HTML, MySQL and Jquery. Experienced professionals on the other hand had more opportunities in Bangalore and southern parts of the country and skills in python, java script, SQL, Agile and Linux were favoured. Both these regions are the top IT and software hubs in India and are top cities for CS&IT professionals as per the Economic Survey 2021-22 (www.indiabudget.gov.in). Among the two, Bangalore has highly impactful software and IT companies as well as startups with large venture capital investments. Job advertisement for Bangalore, Delhi NCR and Mumbai had high number of vacancies whereas high number of applications were received in Bangalore perhaps because of the better perception of the companies in the city or due to the availability of a larger pool of skilled individuals in the area.

The most sought-after skills among freshers and experienced professionals varied between web development languages (javascript, html5, html, php) and general-purpose programming (python), project management (Agile) etc. This could be indicative of the type of responsibilities and job profile assigned to freshers vis-à-vis experienced professionals who were also observed to receive higher average salaries (Indian Rupees 1.28 million per annum). The recommendations extracted for additional skills relevant to each of the top five identified skills are shown in respective section. Lastly, Python and SQL had highest support value in case of job advertisements which received highest number of applications showing that these skills were available with more number of individuals. On the other hand, Java script and Java had highest support values in case of jobs which advertised high number of vacancies indicating that skill development in these areas can slow down in favour of other emerging areas. The jobs in AI & ML are much lesser in number indicating that these skillsets are yet to emerge as areas with large volume of jobs.

## 6. Conclusion

This study has presented a systematic approach for the analysis of skill requirements in specific sectors of an economy using data collected from primary source, namely job advertisements on online job portals. It introduces the step of affinity propagation-based clustering which is an improvement over the methods of expert elicitation used by previous studies. This reduces the time required as well as the number of individuals involved in the analysis. We have conducted a case study of job advertisements for computer science and affiliated industries in India using this approach and identified different aspects of the job market. The skill clusters obtained in two scenarios provide recommendations of skills which can be focused upon by prospective applicants, academic organisations, training institutions, hiring firms etc. This type of analysis may help individuals/ professionals already possessing/ practicing antecedent skills to identify other relevant skills to learn and compete with current demand in the ever-changing job market. Also, academic/ professional development institutions may use this analysis to update their course syllabus to increase the job market readiness of their tutees. It should however be noted that the analysis conducted using this approach will be limited to the areas covered by source website for retrieval of job advertisement data and in that way the findings are as good as the quality of the advertorial data.

**Compliance with Ethical Standards-**
**Competing interest:** All authors certify that they have no affiliations with or involvement in any organization or entity with any financial interest or non-financial interest in the subject matter or materials discussed in this manuscript.